\documentclass[letterpaper,10pt]{article}
\usepackage{osameet2}
\usepackage{graphicx}
\usepackage[cmex10]{amsmath}
\usepackage{amssymb}
\usepackage{amsfonts}
\usepackage{cite} 
\usepackage[dvips,colorlinks=true,bookmarks=false,citecolor=blue,urlcolor=blue]{hyperref}
\begin{document}
\title{Joint Demapping and Decoding for\\ DQPSK Optical Coherent
  Receivers}
\author{Mario A. Castrillon, Damian A. Morero, and Mario R. Hueda}
\address{Digital Communications Research Laboratory - National University of Cordoba - CONICET\\
Av. Velez Sarsfield 1611 - Cordoba (X5016GCA) - Argentina}
\email{acastrillon, dmorero, mhueda@efn.uncor.edu}
\begin{abstract}
  We present a low-complexity joint demapper-decoder scheme for
  coherent optical receivers with DQPSK modulation. The new technique
  reduces to 0.7dB the gap between QPSK and DQPSK in 100Gb/s coherent
  optical systems.
\end{abstract}
\section{Introduction}
Coherent detection based receivers with electronic dispersion
compensation (EDC) are being considered for next generation optical
transport networks (OTN) \cite{1378465}. Quadrature
phase shift keying (QPSK) modulation is the leading candidate for
40Gb/s and 100Gb/s OTN. However, QPSK may suffer from $\pm\pi/2$ phase
jumps or \emph{cycle slips} (CS's), which are induced by phase noise.
These CS's lead to catastrophic bit errors that cannot be corrected
with forward error correction (FEC) codes. The use differential QPSK
(DQPSK) modulation avoids the CS problem at the expense of some
performance degradation.

\begin{figure}[t]
\centering
\includegraphics[width=50mm]{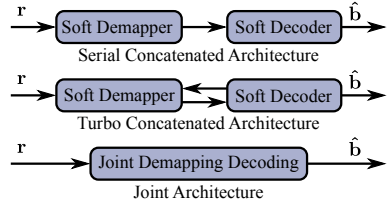}
\caption{Decoder architectures}
\label{fig:fig1}
\end{figure}

Powerful FEC codes with iterative soft-decoding are required for $\ge$
100 Gb/s coherent optical transmission systems. In DQPSK receivers, a
\emph{demapper} block is used to provide soft information to the
iterative channel decoder. Traditional low complexity demappers are
based on a \emph{serial concatenated architecture} (SCA), as depicted
in Fig.\ref{fig:fig1} ~\cite{5621498}.  However, as we
shall show later, the performance of this suboptimal approach with DPQSK modulation is around 1.5 dB worse than that achieved with QPSK \cite{5875249}. This performance degradation can
be combated by using a \textit{turbo concatenated architecture} (TCA)
 or a \textit{joint architecture} (JA)  (see Fig.\ref{fig:fig1}). The use
of these iterative decoding techniques based on the Bahl, Cocke,
Jelinek, and Raviv (BCJR) algorithm, has been reported in past
literature \cite{771340}. Unfortunately, the high
complexity of BCJR-based iterative decoders makes prohibitive their
implementation in multigigabit per second commercial optical receivers.

This paper presents a low complexity \textit{joint demapper and
  decoder} (JDD) architecture for coherent DQPSK modulation. In order
to reduce complexity, the proposed JDD is built upon the \textit{sum
  product algorithm} (SPA) ~\cite{910595}. Although
our JDD is general, we consider here its application with low density
parity check (LDPC) FEC codes. The new JDD extends the \textit{Factor
  Graph} (FG) of the LDPC code by including the minimum set of
additional factor and variable nodes to represent the statistical
relationship between \emph{blocks} of coded bits and received signals.
This way, the accuracy of the decoding process of bit blocks can be
improved at every iteration over a joint factor graph that includes
both the demapper and channel decoding functions. We analyze the JDD
using DQPSK modulation with Gray mapping and an LDPC code with 20\%
overhead (OH) with net-effective coding gain (NCG) of 11.3~dB at a
bit-error-rate of $10^{-15}$ \cite{6133616}. Simulation results
show that the new JDD improves significantly the performance of DQPSK,
providing a 0.6 dB gain over the traditional SCA
~\cite{5621498}.
\section{System Model}
\begin{figure}
\centering
\includegraphics[width=110mm]{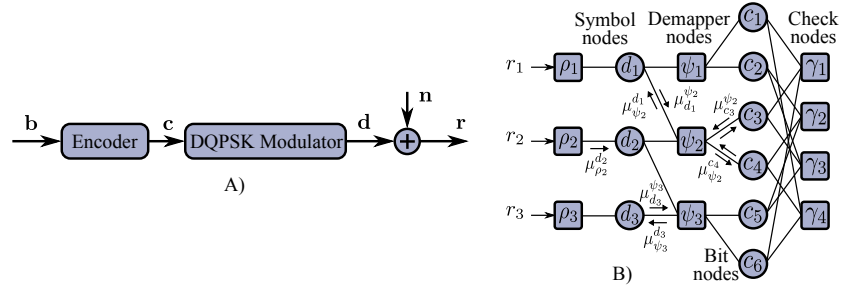}
\caption{A) Transmit system model. - B) Proposed JDD Factor Graph}
\label{fig:fig2}
\end{figure}

Figure \ref{fig:fig2}-(A) shows the transmit system. The information bits
$b_k$ are grouped into blocks of $K$-bits, i.e., $\mathbf{b} \in \{0,1\}^K$. Each
data block $\mathbf{b}$ is encoded with an LDPC code obtaining an $N$-bit block
$\mathbf{c} \in \{0,1\}^N$ where $N$ is even. Codewords
$\mathbf{c}$ are mapped into a sequence of $N/2$ QPSK complex symbols
$\mathbf{s}$, with $\mathbf{s} \in \{1,i,-1,-i\}^{N/2}$. Finally, the components of the transmitted
symbol block $\mathbf{d}$ is computed using differential modulation,
i.e, $d_k = d_{k-1} \cdot s_k$, where $k=1,\ldots,N/2$ and
$d_0=1$. The discrete-time baseband receiver signal is given by
\begin{equation}
r_k = d_k + n_k
\end{equation}
where $k=1,\ldots,N/2$ and $n_k$ are independent identically
distributed (iid) complex Gaussian random variables with zero mean and
variance $N_0$.
\section{Joint Demapper Decoder}
We derive a soft-input soft-output (SISO) joint demapper decoder by
applying the SPA over the FG of the joint a posteriori probability (APP) mass function
$P(\mathbf{c} \mid \mathbf{r})$ of the coded bits given the received
symbols. Based on the Bayes rule, we get $P(\mathbf{c} \mid
\mathbf{r}) = f(\mathbf{r} \mid
\mathbf{c})P(\mathbf{c})/f(\mathbf{r})$. Term $f(\mathbf{r})$ can be
neglected for detection, thus the soft decision can be performed over
$f(\mathbf{r} \mid \mathbf{c})P(\mathbf{c})$. Since
$\mathbf{c}\rightarrow \mathbf{s} \rightarrow \mathbf{d} \rightarrow
\mathbf{r}$ is a Markov chain, the probability density function
$f(\mathbf{r} \mid \mathbf{c})P(\mathbf{c})$ can be expressed as
$f(\mathbf{r} \mid \mathbf{c}) = f(\mathbf{r} \mid \mathbf{d}) \cdot
P(\mathbf{d} \mid \mathbf{s}) \cdot P(\mathbf{s} \mid \mathbf{c})
\cdot P(\mathbf{c})$.  Terms $P(\mathbf{d} \mid \mathbf{s}) \cdot
P(\mathbf{s} \mid \mathbf{c})$ are grouped into one term $P(\mathbf{d}
\mid \mathbf{c})$.  This simplification avoids the implementation of
the variable nodes related the symbols $\mathbf{s}$ and it reduces
the implementation complexity of the SPA with no performance
degradation. Finally, the symbol/bit level factorization of
$f(\mathbf{r} \mid \mathbf{c})P(\mathbf{c})$
is\\[-4mm]
\begin{equation} \label{equ:fg}
f(\mathbf{r} \mid \mathbf{c})P(\mathbf{c}) =
\displaystyle\prod_{k=1}^{N/2} f(r_k \mid d_k)
\displaystyle\prod_{k=1}^{N/2} I(d_k \mid c_{2k-1},c_{2k})
\displaystyle\prod_{i=1}^{M} Q(\mathbf{c}_{(i)}),
\end{equation}
where $\mathbf{c}_{(i)} = \{c_j\in \mathbf{c}\;|\; H_{i,j}=1\}$ are
the coded bits related to the $i$-th parity equation,
$Q(\mathbf{c}_{(i)})=1$ if the bits in $\mathbf{c}_{(i)}$ have even
parity or zero else where and $H$ is the parity check matrix of the
code whose dimensions are $M\times N$, and $I(\cdot)$ is an indicator function.

Figure \ref{fig:fig2}-(B) shows the FG of $f(\mathbf{r} \mid \mathbf{c})P(\mathbf{c})$
according to eq.~\ref{equ:fg}. The messages from node $x$ to node $y$
are denoted by the vector $\mu_{x}^{y}$. The messages between nodes
$c$ and $\gamma$ are computed using the standard SPA for LDPC codes. 
The messages between
nodes $c$ and $\psi$ are 2 dimensional vectors whose components
$\mu_{x}^{y}(b)$ for $b\in\{0,1\}$ represent bit probabilities. The
messages between nodes $\psi$, $d$ and $\rho$ are 4 dimensional
vectors whose components $\mu_{x}^{y}(e^{j\phi})$ for
$\phi\in\{0,\frac{\pi}{2},\pi,\frac{3\pi}{2}\}$ represent symbol
probabilities. The notation $x\langle y \rangle$ represents the two cases of equation, one case using $x$ and other case using $y$. These messages can be computed as:\\[-1mm]
\begin{equation}\label{eq:d2phi}
\mu_{d_k}^{\psi _k\langle \psi _{k+1} \rangle}(e^{j\phi}) = 
\frac{\mu_{\rho_k}^{d_k}(e^{j\phi}) \cdot \mu_{\psi _{k+1}\langle \psi _k \rangle}^{d_k}(e^{j\phi})}
     {\displaystyle\sum_{n=0}^{3} \mu_{\rho_k}^{d_k}\left(e^{j\frac{n\pi}{2}}\right) \cdot \mu_{\psi _{k+1}\langle \psi _k \rangle}^{d_k}\left(e^{j\frac{n\pi}{2}}\right)}
\end{equation}
\begin{equation}
\mu_{\psi _k}^{d_k\langle d_{k-1} \rangle}(e^{j\phi}) = \displaystyle \sum_{n=0}^{1} \sum_{m=0}^{1} \mu_{c_{2k}}^{\psi _k}(n) \cdot \mu_{c_{2k-1}}^{\psi _k}(m) \cdot \mu_{d_{k-1}\langle d_{k} \rangle}^{\psi _k}\left(e^{j\left[\phi -\langle + \rangle\Delta(n,m)\right]}\right)
\end{equation}
\begin{equation}
\mu_{\psi _k}^{c_{2k}\langle c_{2k-1} \rangle}(n\langle m \rangle) = \displaystyle \sum_{m\langle n \rangle=0}^{1} \sum_{l=0}^{3} \mu_{d_{k-1}}^{\psi _k}\left(e^{j\frac{\pi}{2}l}\right)  \cdot \mu_{d_k}^{\psi _k}\left(e^{j\left[ \Delta(n,m)-\frac{\pi}{2}l\right]}\right) \cdot \mu _{c_{2k-1}\langle c_{2k} \rangle}^{\psi _k}( m\langle n \rangle) 
\end{equation}
\begin{equation}
\mu_{\rho _k}^{d_k}(e^{j\phi}) = \frac{1}{\pi N_0} \exp{\left( \frac{-\lVert r_k - e^{j\phi} \rVert ^2  }{N_0}\right) } 
\end{equation}
where $\Delta(n,m) = \frac{(m-n+2mn)\pi}{2}$.
\section{Numerical Results}
\begin{figure}
\centering \includegraphics[width=80mm]{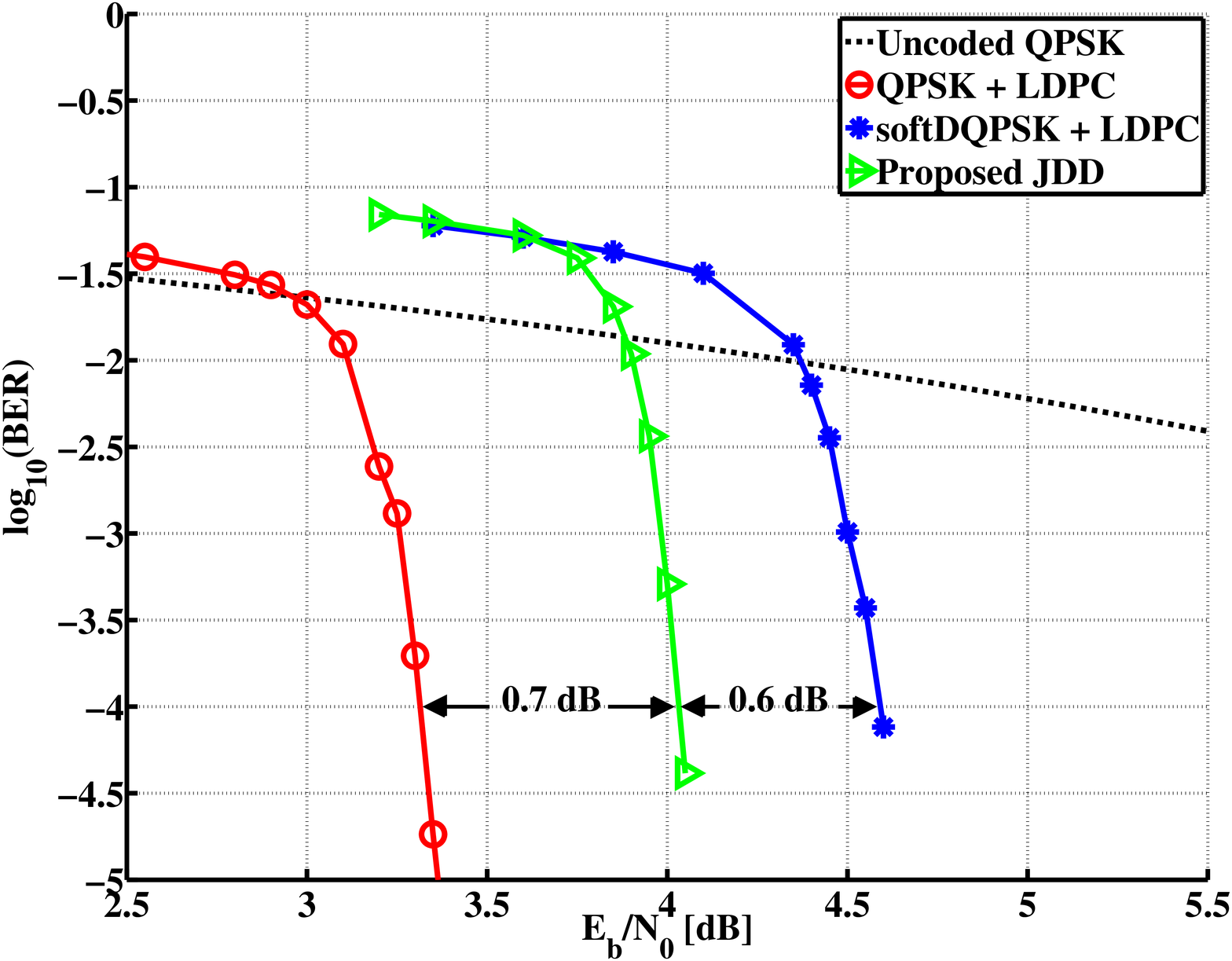}
\caption{Performance of DQPSK and QPSK schemes.}
\label{fig:fig3}
\end{figure}

Figure~\ref{fig:fig3} shows the BER versus the signal-to-noise ratio
per bit ($E_b/N_0$) for the proposed JDD with DQPSK modulation. An
LDPC code of length 24576 bits and rate 0.8334 is used. The number of
iterations in the FG is set to 20. Perfect knowledge of the noise
power $N_0$ is assumed at the receiver size. The performance of an SCA
based on the soft DQPSK demapper defined by eqs. (4) and (5)
in~\cite{5621498}, as well as the performance with QPSK
modulation and optimal demapping, are also presented for comparison
proposes. At least, 1000 errors were counted at each simulation point.
Note that JDD provides a 0.6 dB gain at BER=$10^{-4}$ respect to SCA.
Furthermore, note that the gap between QPSK and DQPSK is reduced to
0.7~dB.
\section{Conclusions}
A new low complexity JDD scheme for coherent DQPSK modulation has been
presented. An SPA-based demapper has been developed to reduce
implementation complexity. Numerical results have shown that JDD
outperforms traditional SCA. This good tradeoff between complexity and
performance makes the proposed JDD an excellent alternative to improve
the performance of DQPSK modulation in next generation optical
transport networks.
\def\refname{References}
\bibliographystyle{osajnl}
\bibliography{paper_arxiv}
\end{document}